\pdfoutput=1
\documentclass[runningheads]{llncs}
\usepackage[T1]{fontenc}
\usepackage{graphicx}
\usepackage{amsmath}
\usepackage{multirow}
\usepackage{booktabs}
\usepackage{makecell}
\usepackage[figuresright]{rotating}
\usepackage{color,xcolor}
\usepackage{subfig}
\usepackage{pifont}
\usepackage{mathrsfs}
\usepackage{multirow}
\usepackage[colorlinks,
            linkcolor=black,
            anchorcolor=black,
            citecolor=black]{hyperref}

\usepackage[misc]{ifsym}

\begin{document}

\title{Practical and General Backdoor Attacks against Vertical Federated Learning}
\author{Yuexin Xuan\inst{1,2}\orcidID{0000-0001-7887-2309} \and
Xiaojun Chen\inst{1,2} \textsuperscript{(\Letter)} \and
Zhendong Zhao\inst{1,2} \and
Bisheng Tang\inst{1,2} \and
Ye Dong\inst{1,2}
}
\authorrunning{Yuexin Xuan et al.}
\institute{School of Cyber Security,University of Chinese Academy of Sciences, Beijing, China \and
Institute of Information Engineering,Chinese Academy of Sciences, Beijing, China
\email{\{xuanyuexin, chenxiaojun, zhaozhendong, tangbisheng, dongye\}@iie.ac.cn}}
\maketitle              % typeset the header of the contribution
\begin{abstract}
Federated learning (FL), which aims to facilitate data collaboration across multiple organizations without exposing data privacy, encounters potential security risks. 
One serious threat is backdoor attacks, where an attacker injects a specific trigger into the training dataset to manipulate the model's prediction.
Most existing FL backdoor attacks are based on horizontal federated learning (HFL), where the data owned by different parties have the same features.
However, compared to HFL, backdoor attacks on vertical federated learning (VFL), where each party only holds a disjoint subset of features and the labels are only owned by one party, are rarely studied.
The main challenge of this attack is to allow an attacker without access to the data labels, to perform an effective attack.
To this end, we propose BadVFL, a novel and practical approach to inject backdoor triggers into victim models without label information.
BadVFL mainly consists of two key steps.
First, to address the challenge of attackers having no knowledge of labels, we introduce a SDD module that can trace data categories based on gradients.
Second, we propose a SDP module that can improve the attack's effectiveness by enhancing the decision dependency between the trigger and attack target.
Extensive experiments show that BadVFL supports diverse datasets and models, and achieves over 93\% attack success rate with only 1\% poisoning rate.

\keywords{Vertical Federated Learning  \and Backdoor Attacks.}
\end{abstract}
\section{Introduction}
Federated Learning (FL), as a promising distributed learning paradigm, 
enables multiple participants to collaboratively train a global model without exposing their private local data. Therefore, it attracts a surge of attention and has been widely applied in many privacy-critical fields like credit risk prediction \cite{kairouz2021advances,credit_risk}, medical diagnosis \cite{medical_dignoise,Kaissis2020SecurePA}, etc.

However, recent works have shown that such promising paradigm encounters severe security threats \cite{liu2020backdoor,jin2021cafe,conf/ndss/ShejwalkarH21,fu2022label,conf/icml/ZhangPSYMMR022}, which significantly hinders its deployment in safety-critical areas. One serious threat to FL is backdoor attacks, where attackers poison partial training data of the victim model to mislead any data with the trigger to a target label, while preserving the model's utility on clean data. It is vital to ensure the security of FL before deployment, as the potential attacks may cause serious threats to the users. For instance, applying a backdoored FL model to the loan risk prediction area, which predicts any users with the trigger as low risk, may lead to huge economic losses.

\begin{figure}[t]
	\centering 
    \includegraphics[width=0.9\textwidth]{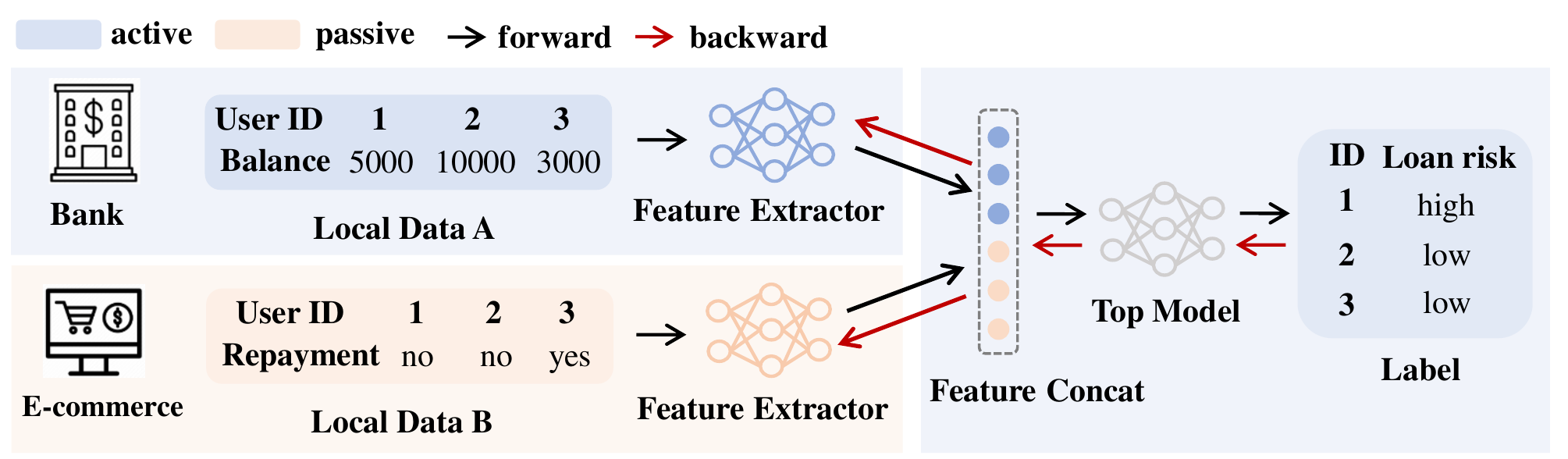}
    	\caption{An example of VFL system. A bank (active party) with account balance features aims to train a more precise model for loan risk analysis by cooperating with an e-commerce company (passive party) holding repayment features. }
	\label{vfl_system}
\end{figure}

FL can be classified into two main categories: horizontal federated learning (HFL) and vertical federated learning (VFL). In HFL, samples sharing the same features are distributed among different participants,
e.g., two regional banks which have different clients but similar businesses like average monthly deposit and account balance to jointly train a model for financial product recommendations.
In VFL, data owned by different parties share the same sample IDs but disjoint features. e.g., a bank with account balance information wants to get a more precise model for loan risk analysis by cooperating with an e-commerce company owning repayment information.
Recent literature has thoroughly analyzed the backdoor attacks and defenses in HFL \cite{xie2020dba,ozdayi2021defending,conf/acsac/XuWKLP22,conf/ndss/RiegerNMS22,conf/uss/NguyenRCYMFMMMZ22}. 
However, the backdoor threats in VFL are rarely explored, despite their increasing relevance in cross-enterprise collaboration. To this end, in this paper, we explore a new backdoor threat in VFL scenario. 

Figure \ref{vfl_system} illustrates the architecture of the VFL system. In VFL, only one party (known as the ``active party") possesses the labels and partial data features, while the other parties (known as the ``passive parties") only have partial data features.
VFL enables the active party to enrich their data features by cooperating with the passive parties who provide more diverse features.
Specifically, in the case where the attacker is the active party, an intuitive way is to add triggers to the local data from the target class, then put them into the training process to implant the backdoor. However, when the attacker is a passive party, the lack of label information makes it more challenging to perform the attack. To address this issue, one may apply label inference to deduce the labels. But the existing state-of-the-art method \cite{fu2022label} requires many auxiliary labeled data and can only perform label inference after the model training is completed, which is impractical for backdoor injection.
Despite the above challenges, we propose BadVFL to conduct backdoor attacks in VFL. Our approach includes two main components.
First, we introduce a Source Data Detection (SDD) module to trace the data categories based on their gradients in run-time. The core idea of SDD is that data from the same class have similar model updating directions.
Second, we propose a Source Data Perturbation (SDP) scheme to enhance the decision dependency between the trigger and attack target, thereby further improving the attack's effectiveness.

We evaluate BadVFL on four benchmark datasets, namely CIFAR-10, ImageNet, BHI, and IMDB, covering both image and text fields. Several excellent results are captured in the experiments. First, our attack is highly effective and general, achieving over 93\% attack success rate with only 1\% poisoning rate on all datasets, while introducing negligible main accuracy drops. 
Second, BadVFL is insensitive to the selection of target data, making it more stable than existing methods.
Third, we evaluate BadVFL against several defense approaches to verify its robustness.

Our technical contributions are summarized below:

\begin{itemize} 
\item We conduct a systematic investigation of backdoor attacks in VFL systems and propose BadVFL, a more general and practical backdoor framework with stable attack performance. Our analysis reveals serious backdoor risks in VFL systems.

\item We propose the SDD module to trace data categories and the SDP module to enhance the dependency between triggers and attack targets. 

\item We conduct extensive empirical validations to show that our framework achieves start-of-the-art performance in terms of effectiveness, generalization, stability, and robustness against several defense methods.
\end{itemize}

\section{Background and Related Work}
\subsection{Vertical Federated Learning} 
VFL~\cite{yang2019federated,Zhang2021SecureBA,Fu2021VF2BoostVF,ijcai/0001ZZWLWWLWZ22} facilitates multiple parties to collaboratively build a model over the partitioned features with privacy-preserving, as all data remains local inside each party. 
Concretely, the VFL protocol is executed as below: 1) the active party broadcasts the sample ID sequence to passive parties to align the data. 2) Each passive party uploads data feature representations extracted by their local model in a predefined order. 3) The active party concatenates these features and feeds them into the top model to calculate the loss and gradients. 4) The active party updates the top model and sends the gradients of uploaded features to passive parties. 5) The passive parties update their bottom models using the received gradients. Supplement B.1 gives the algorithm of the VFL process.

However, such promising training paradigm has been shown to be vulnerable to security threats, such as backdoor attacks \cite{liu2020backdoor}, label inference attacks \cite{fu2022label}, etc.  
It is crucial to ensure the security of VFL systems before deploying them in real-world applications.

\subsection{Backdoor Attacks} 
Backdoor attacks aim at manipulating the victim models’ behavior on backdoored data while maintaining good performance on clean data. Whenever the trigger is presented in the input instance, the backdoor is activated to induce the model to predict the target label.

Backdoor Attacks are first investigated in CV domain \cite{Liu2018TrojaningAO,shafahi2018poison}. Gu et al. \cite{gu2017badnets} generate the poisoned data by adding a specific pattern on clean samples, e.g., a square, and relabeling them with target label before putting them into the training process.
We formulate the loss function of backdoor attacks as below:
\begin{equation} \label{backdoorloss}
    \mathop{\arg\min} \limits_{\theta} \sum \limits_{(x, y) \in D}  \mathcal{L}(\mathcal{F}(x,\theta),y)+\mathcal{L}(\mathcal{F}(x+\mathcal{T},\theta),y_t),
\end{equation}
where $\mathcal{F}$ is the model with parameters $\theta$, $x$ is the clean data with correct label $y$, $\mathcal{T}$ is the trigger and $y_t$ is the target label. The key to backdoor attacks is to establish a strong link between the trigger and the attack target, which is achieved by the last term in Eq. \ref{backdoorloss}.

Recent studies have explored the backdoor attacks in the HFL scenario, which is more vulnerable due to clients having full control over the local labeled data and the training process, making it easier to submit malicious updates to build up a mapping between the trigger and target label. Xie et al. \cite{xie2020dba} introduce a distributed backdoor attack by decomposing a global trigger into several local triggers and assigning them to different adversarial clients. Bagdasaryan et al. \cite{bagdasaryan20a} explore a model replacement approach by scaling the malicious model updates to replace the global model with the local poisoned one.

However, backdoor attacks in the VFL scenario are rarely explored because the attack achieved by the passive party is more challenging due to the lack of label information. Liu et al. \cite{liu2020backdoor} introduce a gradient replacement (GR) approach by replacing the gradient of local triggered samples with the gradient of the target data (explained in Section 4) when updating the local model. However, GR heavily relies on the selection of the target data and neglects the impact of features owned by other parties in the final classification.

\section{Problem Formulation}

\subsection{Vertical Federated Learning} 
In a VFL system, there are $K$ parties $\{P_k\}_{k=1}^K$, where each party $P_k$ holds partial features and the labels are privately owned by the active party. 
We denote the whole training dataset as $D=\{x_i=(x^1_i, x^2_i, ..., x^K_i), y_i\}^N_{i=1}$, where $x^k_i$ is the feature of $i$th sample located on $k$th party, and $y_i$ is the true label of $i$th sample.
Each party holds a local feature extractor $\mathcal{F}_{\theta_k}$ to transform the local data $x_i^k$ into feature representations. VFL minimizes the following loss function to ensure performance: 
\begin{equation}
    \begin{aligned}
    \mathop{\arg\min} \limits_{\theta} \sum \limits_{(x_i, y_i) \in D} \mathcal{L}^{ce} \left( \mathcal{M}_{\theta_t} (f_i) ,y_i \right) + \Omega(\theta),
    \end{aligned}
\end{equation}
where $f_i = Concat\{ \mathcal{F}_{\theta_1}(x_i^1),...,\mathcal{F}_{\theta_K}(x_i^K)\}$ is the merged feature representation of the $i$th sample, $\mathcal{L}^{ce}$ is the cross-entropy loss and $\mathcal{M}_{\theta_t}$ is the top model,  $\theta_k$ is the parameters of local feature extractor $\mathcal{F}_{\theta_k}$ owned by $P_k$, and $\Omega(\theta)$ is the regularization term to avoid overfitting. 

\subsection{Threat Model}
As stated previously, we assume one of the passive parties with no label information is the adversary. Without loss of generality, we assume $P_K$ is the attacker.

\subsubsection{Attacker's Goal}
The goal of $P_K$ is to establish a strong link between the trigger and the attack target. Whenever the trigger is presented in the input instance, the victim model should predict the target label. Meanwhile, the attacker should ensure the clean data are classified correctly to maintain the model utility. Formally, the attacker optimizes the following objective function: 

\begin{equation}\label{bad_loss}
\begin{aligned}
    \mathop{\arg\min} \limits_{\theta} \sum \limits_{(x_i, y_i) \in D} \mathcal{L}^{ce} \left( \mathcal{M}_{\theta_t} (f_i^{c}),y_i \right) 
     + \mathcal{L}^{ce} \left( \mathcal{M}_{\theta_t} (f_i^{p}),y_t \right) + \Omega(\theta),
\end{aligned}
\end{equation}
where $ f_i^{c} = Concat \{ \mathcal{F}_{\theta_1}(x_i^1),...,\mathcal{F}_{\theta_K}(x_i^K)\}$, $f_i^{p} = Concat \{ \mathcal{F}_{\theta_1}(x_i^1),...,\mathcal{F}_{\theta_K}(x_i^K+ \mathcal{T})\}$ are the clean and poisoned feature representations, respectively. $\mathcal{T}$ is the injected trigger. The first term ensures that the victim model behaves normally on clean data, and the second term achieves the backdoor behavior.
We showcase various backdoored samples in Supplement A.1.

\subsubsection{Attacker's Capability}
We assume $P_K$ strictly follows the VFL protocols: uploading feature representations, receiving gradients, and updating its local model. The data accessible to $P_K$ are: own local data $\{x_i^K\}_{i=1}^N$ and the corresponding gradients $\{g_i^K\}_{i=1}^N$ returned from the active party. Moreover, $P_K$ has no knowledge of the data, the model, and any intermediate information owned by other parties. The adversary cannot interfere with the normal interactions between the active party and other passive parties. 

\section{Backdoor Attacks in VFL}

In this section, we present a detailed explanation of how BadVFL can achieve backdoor attacks in the VFL systems.

The key to successful backdoor attacks is associating a pre-defined trigger with the target label. One intuitive method is adding the trigger into the data from the target class to link the trigger with the attack target. However, $P_K$ without label information does not know which data comes from the target class. To address this issue, we design the Source Data Detection (SDD) module, which can infer data categories based on their gradients in run-time. Moreover, to further improve the attack's effectiveness, we propose the Source Data Perturbation (SDP) scheme, which enhances the decision-dependency between the trigger and the attack target.
Figure \ref{framework} shows the steps involved in BadVFL. And the detailed algorithm of BadVFL process is given in Supplement B.2.

\begin{figure}[t]
	\centering
    \includegraphics[width=0.9\textwidth]{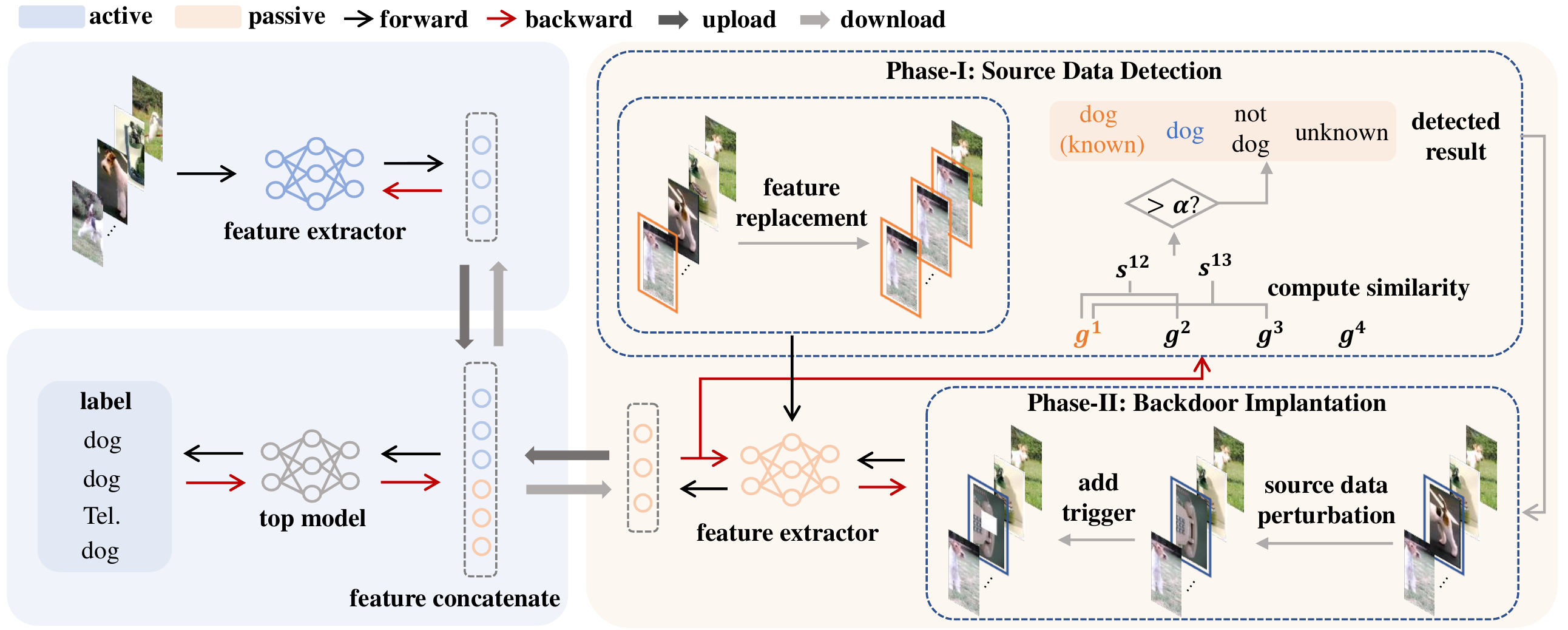}
    	\caption{The framework of BadVFL, which contains two main modules: Source Data Detection and Backdoor Implantation. The former aims to detect the source data (marked in blue) based on the target data (marked in orange).
        }
	\label{framework}
\end{figure}

\begin{definition}
\textbf{(Target Data)} \textit{Target data is a prior knowledge which comes from the target class known by the attacker.}
\end{definition}

\begin{definition}
\textbf{(Source Data)} \textit{Source data are obtained by SDD module used for data poisoning which have the same label with target data.}
\end{definition}

\subsection{Source Data Detection}
We assume $P_K$ only knows one target data denoted as $x_{t}^K$, where $t$ is the sample ID of target data. (Note that data should be uploaded to the top model in a pre-defined order.)
Intuitively, the key insight of SDD is that data from the same class will have similar model updating directions (a.k.a., gradients). With known $x_{t}^K$, $P_K$ normally participates in the VFL training process until first getting $g_t^K$. Then $P_K$ runs SDD to infer which data comes from the target class by computing the similarity between $g_t^K$ and the gradients of other data. In detail, the process involves two main steps: \textit{feature replacement} and \textit{similarity computation}.

\subsubsection{Feature Replacement} We randomly select $n$ samples per batch (denoted as $x_{nset}^K$, $|nset|=n$) to detect whether they are from the target class. To increase the gradient similarity between $x_{t}^K$ and the data from the target class in $x_{nset}^K$, we replace $x_{j}^K$ with $x_t^K$ (for all $j\in nset$), so the only difference between $x_t$ and $x_{nset}$ is the data held by the other clients. Then we upload the replaced data to the top model.

\subsubsection{Similarity Computation} After obtaining the returned gradients of $x_{nset}^K$ (denoted as $g_{nset}^K$), we compute the cosine similarity between $g_{nset}^K$ and $g_t^K$. (Note that $g_t^K$ is updated when $x_t^K$ is uploaded.)

\begin{equation}
cos(g_t^K,g_j^K) = \frac {\left \langle g_t^K, g_j^K \right \rangle}{\left \|g_t^K \right\|_2 \left \| g_j^K \right \|_2}, \quad for \ all \ j \in nset.
\end{equation}
Intuitively, the higher the similarity, the more likely they have the same label.
To illustrate how this works, we consider a simple example where the top model consists of only one linear layer. 
\begin{equation}
    \left (
    \begin{matrix}
    W_{11} & \cdots & W_{1d} \\\
    \vdots & \ddots & \vdots \\\
    W_{C1} & \cdots & W_{Cd} 
    \end{matrix}\right ) 
    \left(
    \begin{matrix}
    \mathcal{F}_{\theta_1}^{T}(x_i^1)\\\
    \vdots \\\
    \mathcal{F}_{\theta_K}^{T}(x_i^K)
    \end{matrix} \right ) = 
    \left (
    \begin{matrix}
    O_1 \\\
    \vdots \\\
    O_C
    \end{matrix}
    \right),
\end{equation}
where $W$ is the top model parameters, $O$ is the output of sample $i$, and $C$ is the number of classes for classification. Here we assume the true label $y_i$ of sample $i$ is class $c$, where $c \in [1, C]$. As we can see, the gradient of the cross-entropy loss w.r.t. the feature representation of sample $i$ is: 
\begin{equation}
   \frac {\partial \mathcal{L}^{ce}(f_i; W, y_i)}{\partial f_i} = 
   \left (
   \begin{matrix}
   W_{c1}, \cdots, W_{cd}
   \end{matrix}
   \right) = W_c.
\end{equation}

Therefore, the gradients received by $P_K$ have the following properties: data from the same class will return similar gradients, resulting in a high positive cosine similarity. While samples from different classes will return dissimilar gradients, resulting in a low cosine similarity. 

Finally, we set a threshold $\alpha_{thre}$. If the similarity between $g_{j}^K$ (for $j \in nset$) and $g_t^K$ is higher than $\alpha_{thre}$, we consider $x_j^K$ to be from the target class, named source data. We terminate the SDD process until enough source data are found or all training data have been considered. Compared with the state-of-the-art label inference method \cite{fu2022label} which requires many auxiliary labeled data,
our SDD is more practical.

\subsection{Backdoor Implantation} 
Based on the above steps, we have already inferred the source data which are from the target class. Our next task is to associate a pre-defined trigger with the attack target. One intuitive method is directly adding the trigger to the source data and putting them into the training process. However, the top model may still learn the mapping between the background clean features of the source data and the target label, causing the failure of backdoor injection. 
Therefore, we propose a Source Data Perturbation (SDP) module to further enhance the decision-dependency between the trigger and the attack target.

\subsubsection{Source Data Perturbation.} A successful backdoored model should give the target prediction as long as the malicious trigger is present, despite the existence of the background clean features. To make the trigger a higher priority than the other clean features in the decision-making phase, we attempt to replace the source data with the data randomly selected from the same batch. 
In this way, the source data will contain clean features of different classes. Then we add the trigger on the perturbed source data and put them into the training process to achieve backdoor injection.

There are mainly two reasons why SDP module enhances the decision-dependency between the trigger and attack target: 
(a) The source data inferred by SDD have the same label (attack target). And their features are replaced by randomly selected data after the SDP module. This makes the source data equipped with different features but the same label, causing the model more difficult to learn from these features. (b) After the SDP process, we add the same trigger to the source data, thus they have the same trigger and the same target label, which makes the model more likely to establish the decision-dependency between the trigger and attack target.

\section{Experiments}
\subsection{Experiment Setup}
As most real VFL systems consist of two parties \cite{model_inversion,chen2021homomorphic,fu2022label}, for the rest of this paper, we construct and evaluate BadVFL under a two-party scenario. The attacks in multi-party settings are given in Supplement C.1.

\subsubsection{Datasets \& Networks.} We evaluate BadVFL on the following datasets: CIFAR-10~\cite{krizhevsky2009learning}, ImageNet~\cite{russakovsky2015imagenet}, Breast Histopathology Images (BHI)~\cite{BHI_dataset}, and IMDB~\cite{maas2011learning}. The first three are image datasets, and IMDB is a text dataset. We describe
the datasets in detail in Supplement A.2.
To make these datasets suitable for the VFL scenario, as the common setting in VFL \cite{liu2020backdoor,jin2021cafe,fu2022label}, for CIFAR-10 and ImageNet, we split the data into two parts along the middle line so that each party holds half. For BHI, there are multiple examination image patches per patient, and we distribute the patches of each patient with the same label to each party in a round-robin manner. 
For IMDB, we split each sample (a paragraph for a movie review) into two parts and distribute them to each party. 

We experiment on three classic deep neural networks to get feature representations, namely ResNet18 \cite{he2016deep} for CIFAR-10 and BHI, VGG16 \cite{simonyan2015very} for ImageNet and LSTM \cite{1997Long} for IMDB. As for the top model used for feature combination and classification, following previous works \cite{Hu2019FDMLAC,jin2021cafe}, we adopt a linear combination of these features and then apply a nonlinear transformation (e.g., softmax) to make the prediction. To verify the stability of BadVFL, we also conduct experiments with multi-hidden layers. 

\subsubsection{Implementation Details.} For image datasets, the models are trained by SGD optimizer for 200 epochs. The initial learning rate is $0.01$, multiplied by $0.1$ per $50$ epoch. For text dataset, the optimizer is Adam with an initial learning rate of $0.001$. In all experiments, the poisoning rate $\eta = \frac {|D_{poisoned}|}{|D_{train}|}$ is $1\%$, as the common setting for backdoor attacks \cite{gu2017badnets}. And we set the replacement number $n=5$ and the threshold $\alpha_{thre}=0.6$ in SDD for all datasets.

\subsubsection{Evaluation Metrics.} We adopt Test Accuracy Rate (TAR) and Attack Success Rate (ASR) to evaluate BadVFL performance. Specifically, TAR is the probability that the clean data are classified correctly, measuring the impact of backdoor attacks on the main task. ASR is the probability of predicting the poisoned data as the target label, which measures the attack efficacy.

\subsection{Attack Performance}
\subsubsection{Attack Effectiveness.} For image datasets, the trigger we applied following Gu et al. \cite{gu2017badnets}, which is a white square located in the center of the image. We apply $4\times4$, $20\times20$, and $5\times5$ trigger size for CIFAR-10, ImageNet and BHI, respectively. For IMDB, we insert the word `[START]' into the middle of the sentence as the trigger. 

To ensure our attack remains consistently effective, for each class, we construct BadVFL with randomly selecting 3 different target data from the dataset and get their average as the final result. 
To suppress the effect of non-determinism, all experiments are averaged across multiple runs. 
The results are shown in Table \ref{main_result}. Observe that the triggers are successfully injected as the poisoned models have small TAR difference compared with the benign models and high ASR. Specifically, BadVFL achieves above $93\%$ ASR in all datasets with negligible TAR drops.

\begin{table}[h]
\setlength{\tabcolsep}{5pt}
\centering \caption{The attack performance of BadVFL and GR in four datasets.}\label{main_result}
\begin{tabular}{ccccccccc}
\hline
\multirow{2}{*}{\textbf{Dataset}} & \multirow{2}{*}{\textbf{}} & \multirow{2}{*}{\textbf{\begin{tabular}[c]{@{}c@{}}Benign\\ VFL\end{tabular}}} & \multirow{2}{*}{\textbf{}} & \multicolumn{2}{c}{\textbf{BadVFL}} & \textbf{} & \multicolumn{2}{c}{\textbf{GR}} \\ \cline{5-6} \cline{8-9} 
                                  &                            &                                                                                &                            & \textbf{TAR}     & \textbf{ASR}     & \textbf{} & \textbf{TAR}   & \textbf{ASR}   \\ \hline
\textbf{CIFAR-10}                 &                            & 80.96                                                                          &                            & 80.69            & 94.98            &           & 77.08          & 44.07          \\ \hline
\textbf{ImageNet}                 &                            & 79.63                                                                          &                            & 79.47            & 93.15            &           & 73.01          & 19.21          \\ \hline
\textbf{BHI}                      &                            & 91.90                                                                          &                            & 89.52            & 99.11            &           & 88.45          & 98.93          \\ \hline
\textbf{IMDB}                     &                            & 85.62                                                                          &                            & 85.01            & 98.97            &           & 81.99          & 51.98          \\ \hline
\end{tabular}
\end{table}

\begin{figure}[!h]
	\centering 
    \includegraphics[width=0.8\textwidth]{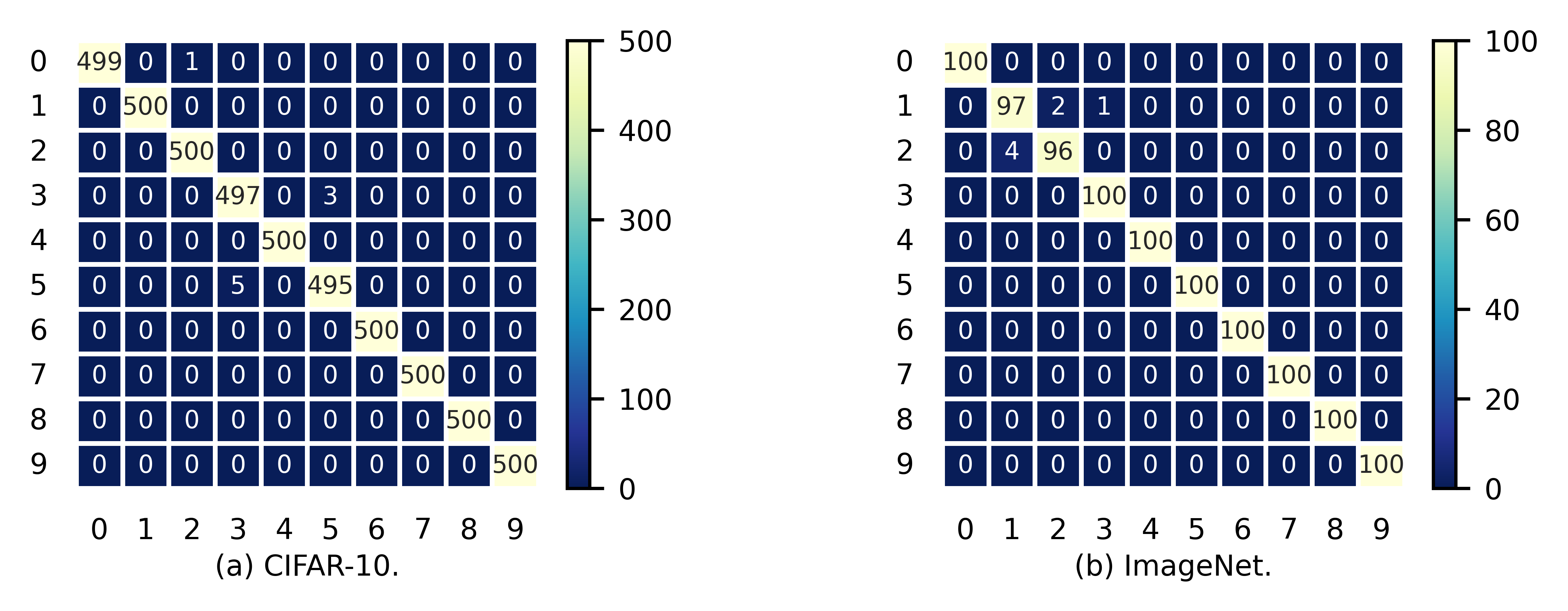}
    	\caption{The number of source data from class $j$ (column) being identified as class $i$ (row) by SDD on CIFAR-10 (left) and ImageNet (right) dataset. The total number of source data is 500 for CIFAR-10 and 100 for ImageNet.}
	\label{correct_count}
\end{figure}

Figure \ref{correct_count} depicts the category distribution of the source data obtained by SDD module for different target class. The color-coded values in row $i$ and column $j$ represent the number of inferred source data from class $j$ when target data from class $i$.
As we can see, for most cases, SDD module can correctly identify the source data which are truly from target class. However, for some target class, such as class $3$ in CIFAR-10, the SDD mistakenly identifies few samples from class $5$ as its source data. This is because the data from class $3$ (cat) have the similar features with data from class $5$ (dog). And few false detection results have little influence on the attack effectiveness.

\subsubsection{Comparison with Gradient Replacement.} We compare BadVFL with the state-of-the-art method GR \cite{liu2020backdoor}, which attacks VFL system by replacing the gradient of local triggered samples with the gradient of the target data when updating the local bottom model.
Table \ref{main_result} shows the comparison results. Apparently, our method outperforms GR on all metrics and achieves a significant boost.
In more details, GR hurts worse on the main task accuracy. This is because the adversary replaces the triggered data's feature with random vectors and sends them to the active party, to prevent the active party establishing a new mapping between the triggered data with their true label. Moreover, the attack performance of GR strongly depends on the selection of target data.

\subsection{Multi-Hidden Layers Performance}

\begin{figure}[h]
	\centering 
    \includegraphics[width=0.75\textwidth]{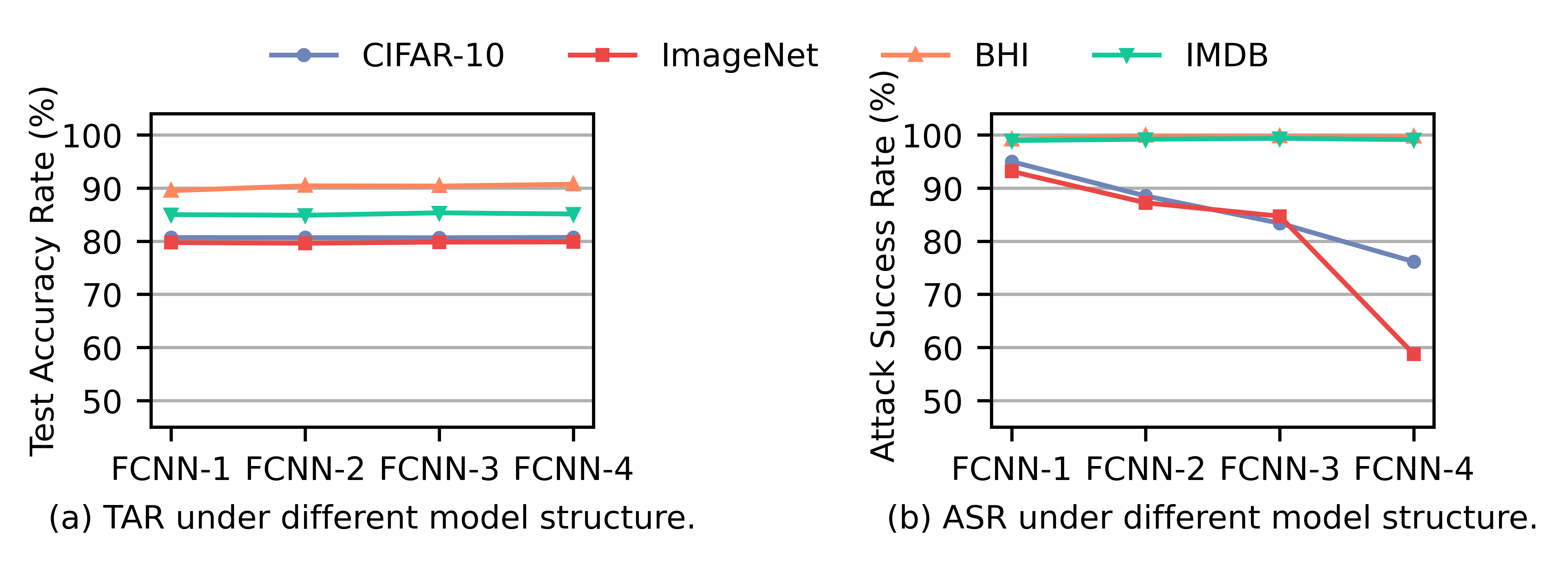}
    	\caption{The performance of BadVFL \textit{w.r.t.} different top model structures.}
	\label{multi_layer}
\end{figure}

To verify the stability of BadVFL with different top model structures, in this section, we show the effectiveness of BadVFL for multi-hidden fully-connected neural networks. Considering the role of top model is feature combination and classification, the structure of it does not need to be complex. As shown in Figure \ref{multi_layer}, we experiment on 1-hidden, 2-hidden and 3-hidden layers with ReLU activation followed by softmax transformation.  

Intuitively, for BHI and IMDB dataset, there is no significant difference with the increasing of model depth. However, for CIFAR-10 and ImageNet dataset, the BadVFL performs worse as the network becomes deeper. Diving to the bottom, the phenomenon is caused by the fact that the more complicated top model structure affects the gradient-based SDD module calculation, leading to wrongly inferred source data which are not from the target class. 

\subsection{Defenses}
To demonstrate how defensive strategies against BadVFL, we conduct BadVFL with noisy gradients and gradient compression, which are commonly used by prior works to train the robust FL systems \cite{hitaj2017deep,fu2022label}.

\subsubsection{Noisy Gradients.} One straightforward attempt to defense BadVFL is adding noise to the exchanged information. We experiment Gaussian noise \cite{NEURIPS2019_60a6c400} with variance from $10^{-5}$ to $10^{-2}$. Because adding noise inevitably affects the gradient similarity calculation, we select the 1\% highest similarity results as the source data in each iteration instead of a fixed threshold. The results are shown in Figure \ref{defense_noise}. As we can see, the BadVFL performance monotonically decreases with the increasing of noise scales.
In details, when variance is 1e-4, the ASR on ImageNet severely deteriorates and TAR drops by nearly 20\%, resulting in a good defense performance but seriously compromising the model utility. For CIFAR-10, setting variance to 1e-3 successfully defends against BadVFL, where the ASR drops to 30\% with negligible TAR drops.
However, in practice, it is non-trivial to figure out an appropriate noise scale that guarantees security while maintains model utility.

\subsubsection{Gradient Compression.} 
\begin{figure}   
  \centering            
  \subfloat[Noisy Gradients.] 
  {
      \label{defense_noise}\includegraphics[width=0.7\textwidth]{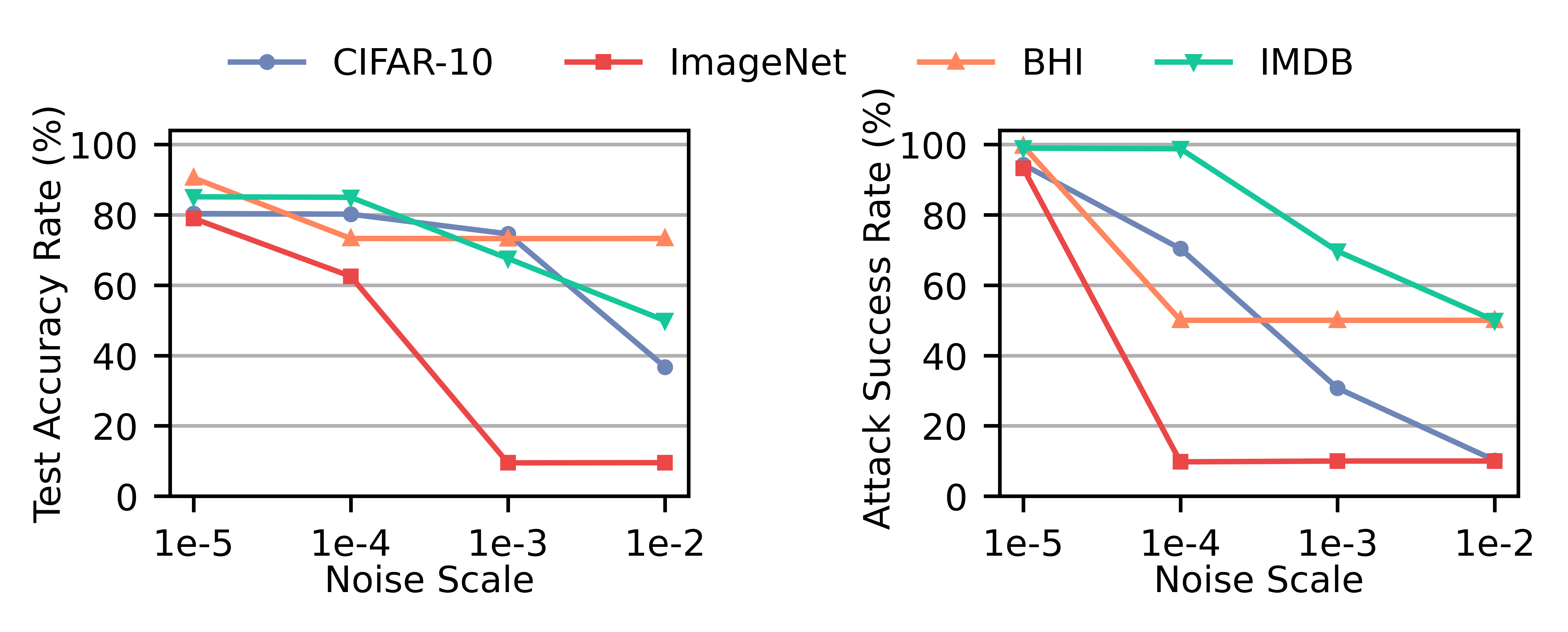}
  } \\
  \subfloat[Gradient Compression.]
  {
      \label{defense_compress}\includegraphics[width=0.7\textwidth]{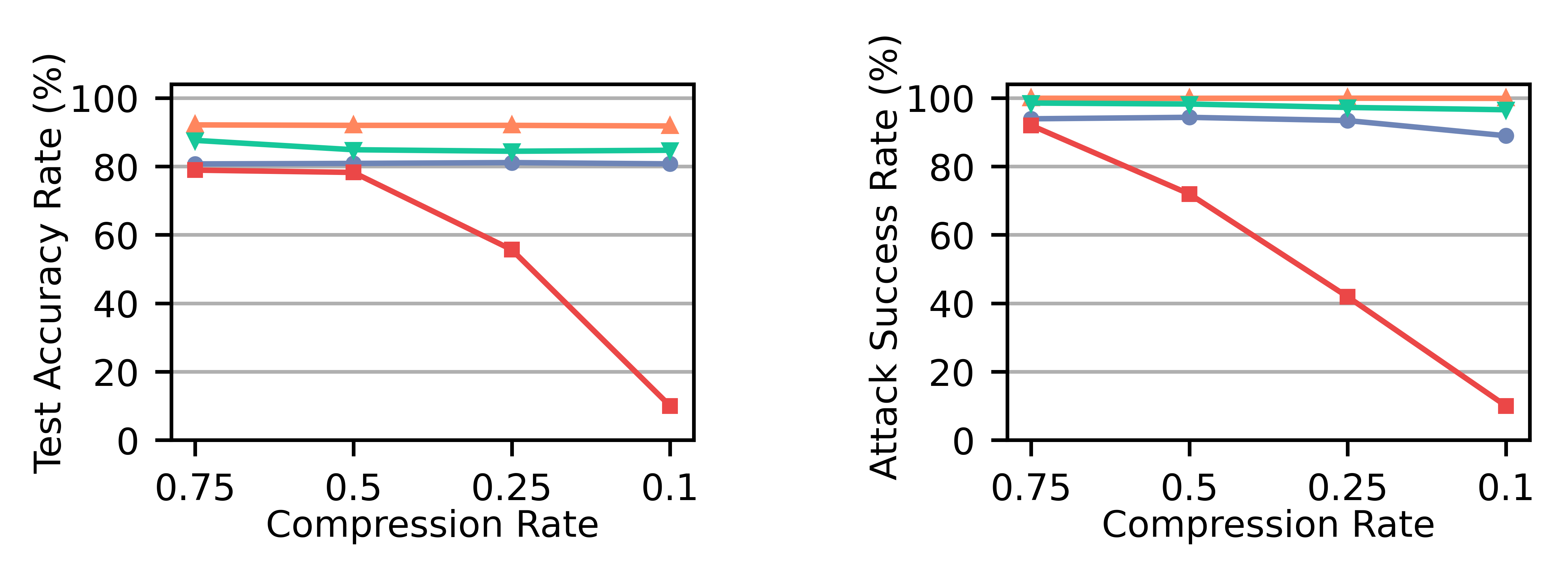}
  }
  \caption{BadVFL performance against different defense strategies on all datasets.}    
  \label{fig:subfig_1}        
\end{figure}

Another effective defense strategy is pruning the gradients with small magnitudes to zero \cite{yujun2017}. We evaluate different level of sparsity from $0.75$ to $0.1$. As shown in Figure \ref{defense_compress}, interestingly, for CIFAR-10, BHI and IMDB, BadVFL maintains considerable high TAR and ASR with the increasing of compression rate. As for ImageNet, the gradient compression can successfully mitigate the backdoor attacks in VFL, but introducing significantly TAR drops and destroying the model utility. 

\subsection{Ablation Study}
\subsubsection{Position of Trigger.} 
\begin{figure}[h]
	\centering 
    \includegraphics[width=0.7\textwidth]{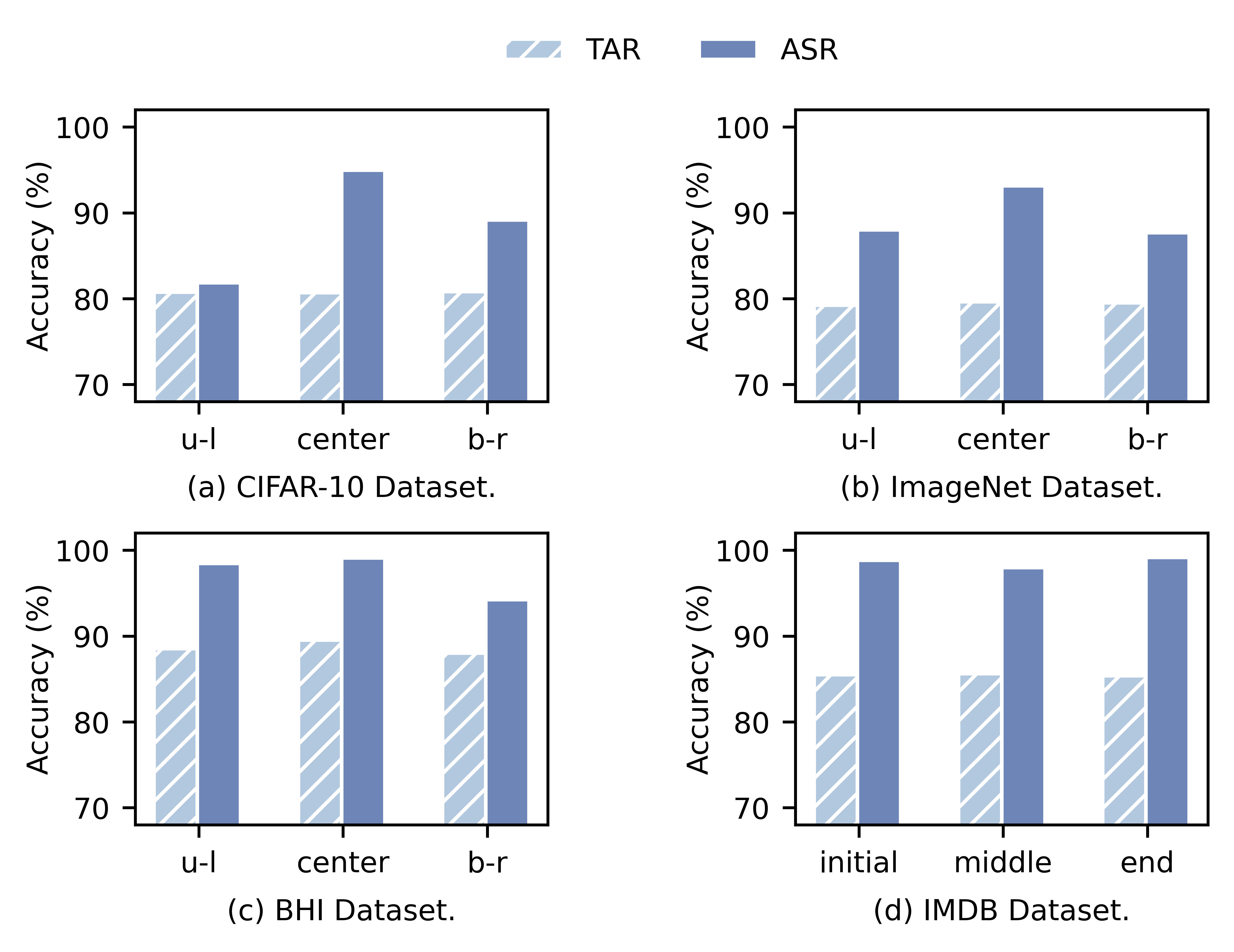}
    	\caption{BadVFL performance \textit{w.r.t} varied trigger positions.}
	\label{trigger_position}
\end{figure}

To further validate the influence of trigger position on attack effectiveness, we plot Figure \ref{trigger_position} to show the BadVFL attack performance with three possible locations. For image classification task, we experiment the trigger located in ``up left" (u-l), ``center", and ``bottom right" (b-r) of the image to analyze the effectiveness. For text classification task, we conduct the experiments with the trigger located in the ``initial", ``middle" and ``end" of the text.

As shown in Figure \ref{trigger_position}, we notice that for image classification task, the center location has a significant advantage over the other two locations, because the center area contributes more to model classification and its original features are blocked. 

For text classification task, the initial and end position have a slight advantage over the middle position, due to the training mode of the Recurrent Neural Networks.
Nevertheless, no matter where the triggers are, BadVFL can always succeed in injecting the backdoor into the model with negligible main accuracy drops.

\subsubsection{Source Data Perturbation.} 
As discussed in Section 4, if we directly add triggers to the source data and put them into training process, the model may not learn the trigger but the clean feature of the source data. Thus, we evaluate the importance of SDP and the results are shown in Table \ref{data_perturb}. Specifically,
we consider the three following cases: (a) without perturbation, (b) replace the source data with data selected from the same batch, and (c) replace the source data with data selected from the whole dataset. 

We observe that there is no significant difference between the cases where the attacker replaces the source data with the data from same batch, or from the whole training dataset. However, when the attacker does not perturb the source data and directly adds the trigger on them, the BadVFL performance drops significantly. This is especially prominent in the case of the CIFAR-10 and ImageNet dataset.

\begin{table}[]
\setlength{\tabcolsep}{4pt}
\centering \caption{BadVFL performance \textit{w.r.t.} different types of source data perturbation.} \label{data_perturb}
\begin{tabular}{ccccccccccccc}
\hline
\multirow{2}{*}{\textbf{Dataset}} &  &  & \multicolumn{2}{c}{\textbf{No Perturb}} &  & \textbf{} & \multicolumn{2}{c}{\textbf{\begin{tabular}[c]{@{}c@{}}Replace from \\ Same Batch\end{tabular}}} &  & \textbf{} & \multicolumn{2}{c}{\textbf{\begin{tabular}[c]{@{}c@{}}Replace from \\ Whole Dataset\end{tabular}}} \\ \cline{4-5} \cline{8-9} \cline{12-13} 
                                  &  &  & \textbf{TAR}       & \textbf{ASR}       &  & \textbf{} & \textbf{TAR}                                   & \textbf{ASR}                                   &  & \textbf{} & \textbf{TAR}                                     & \textbf{ASR}                                    \\ \hline
\textbf{CIFAR-10}                 &  &  & 80.90              & 46.16              &  &           & 80.69                                          & 94.98                                          &  &           & 80.67                                            & 94.80                                           \\
\textbf{ImageNet}                 &  &  & 79.38              & 44.13              &  &           & 79.47                                          & 93.15                                          &  &           & 79.03                                            & 93.84                                           \\
\textbf{BHI}                      &  &  & 87.82              & 87.93              &  &           & 89.52                                          & 99.11                                          &  &           & 90.25                                            & 99.32                                           \\
\textbf{IMDB}                     &  &  & 85.29              & 63.34              &  &           & 85.01                                          & 98.97                                          &  &           & 85.45                                            & 98.33                                           \\ \hline
\end{tabular}
\end{table}

\subsection{Hyperparameter Analysis}
\subsubsection{Influence of injecting rate $\eta$.} We investigate the critical factor $\eta$ which affects the number of poisoned data in the training process. As shown in Figure \ref{param}a, the ASR becomes worse when $\eta > 3\%$. 
This is because the attacker uploads more ``wrong'' features (perturbed source data) as $\eta$ increases, resulting in the top model depends more on other clean features in decision-making.

\subsubsection{Influence of replacement number $n$.} We depict the impact of replacement number $n$ on the BadVFL performance in Figure \ref{param}b. There is no significant different among varied $n$. Moreover, because a small $n$ makes the poisoning process more stealthy, we set $n=5$ as default for all datasets. 

\subsubsection{Influence of threshold $\alpha_{thre}$.} The $\alpha_{thre}$ in the SDD module can control the final source data set of backdoor attacks. The results are shown in Figure \ref{param}c. Specifically, when $\alpha_{thre} < 0.6$, the larger $\alpha_{thre}$ makes BadVFL more effective. This is caused by the fact that the larger $\alpha_{thre}$ can infer more accurate source data which are truly from target class. When $\alpha_{thre} > 0.6$, the BadVFL converges more stably. Hence, we set $\alpha_{thre}=0.6$ for all datasets.

\subsubsection{Influence of trigger size $ts$.} 
\begin{figure}[!h]
	\centering 
    \includegraphics[width=0.7\textwidth]{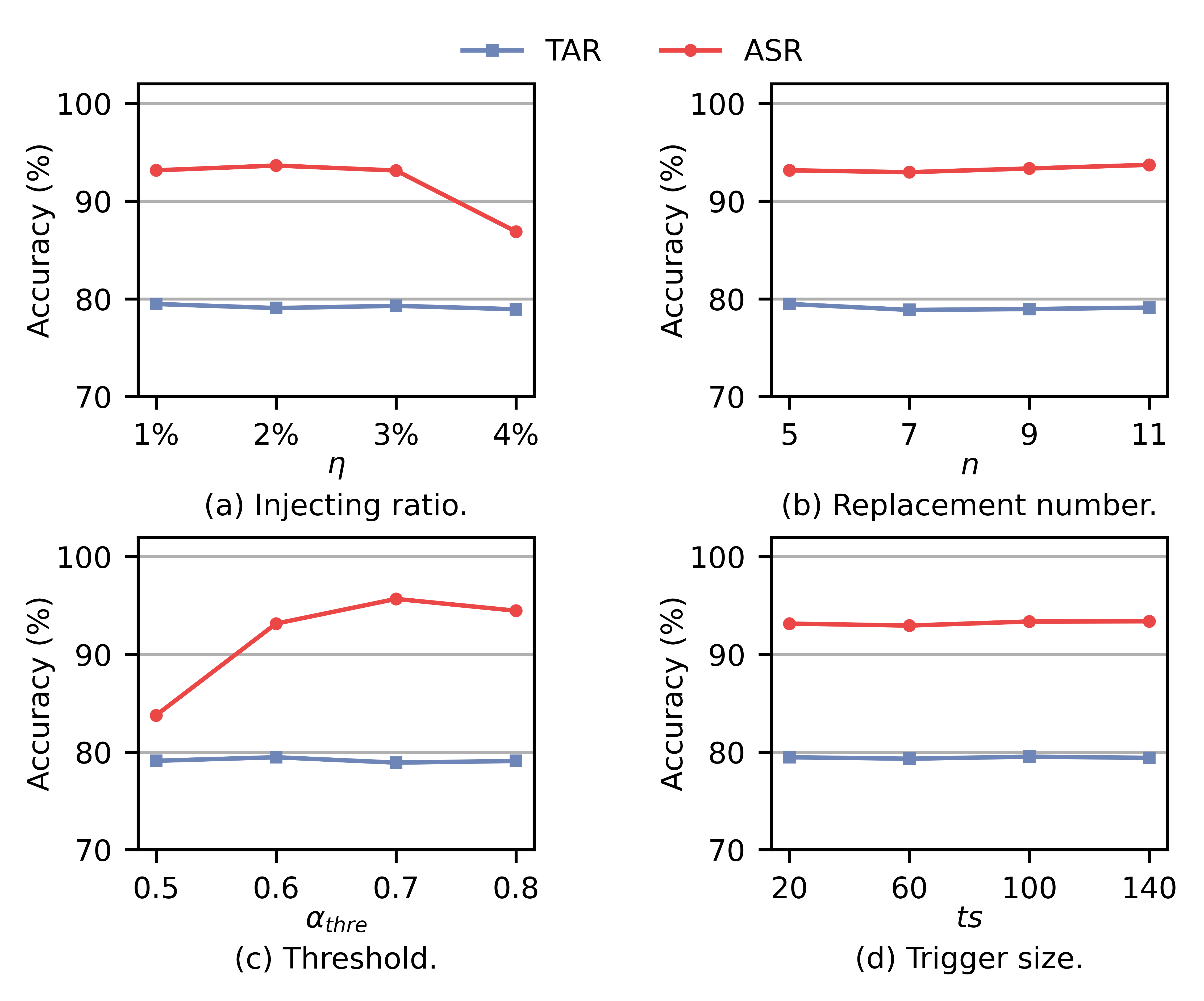}
    	\caption{The performance of BadVFL on ImageNet dataset \textit{w.r.t.} different hyperparameters.}
	\label{param}
\end{figure}

Another critical factor that might affect the BadVFL performance is the size of trigger. As shown in Figure \ref{param}d, we can observe that BadVFL achieves a stable performance among different trigger sizes.

\section{Conclusion}
In this paper, we have demonstrated a new security risk where backdoor attacks can be successfully implanted into VFL systems. We propose BadVFL, 
which outperforms state-of-the-art method remarkably.
It is likely that future defenses will defeat this attacks,
however, we believe that the attacks are a promising direction that (a) expose possible security threats in such promising training paradigm, and (b) enlighten the future work.
Future direction suggested by our work is the counter measures against backdoor attacks under VFL scenario, such as anomaly data detection and backdoor mitigation, etc. However, these might be difficult because the defender only holds partial data features and part of global model.

\section{Acknowledgements} 
This work is supported by the Strategic Priority Research Program of Chinese Academy of Sciences, Grant No. XDC02040400.

% \section{Ethical Statement}
% There are no ethical issues in our work.

\bibliographystyle{splncs04}
\bibliography{mybibliography}

\begin{thebibliography}{10}
\providecommand{\url}[1]{\texttt{#1}}
\providecommand{\urlprefix}{URL }
\providecommand{\doi}[1]{https://doi.org/#1}

\bibitem{bagdasaryan20a}
Bagdasaryan, E., Veit, A., Hua, Y., Estrin, D., Shmatikov, V.: How to backdoor
  federated learning. In: Proceedings of the Twenty Third International
  Conference on Artificial Intelligence and Statistics (2020)

\bibitem{chen2021homomorphic}
Chen, C., Zhou, J., Wang, L., Wu, X., Fang, W., Tan, J., Wang, L., Liu, A.X.,
  Wang, H., Hong, C.: When homomorphic encryption marries secret sharing:
  Secure large-scale sparse logistic regression and applications in risk
  control. In: Proceedings of the 27th ACM SIGKDD Conference on Knowledge
  Discovery \& Data Mining. pp. 2652--2662 (2021)

\bibitem{ijcai/0001ZZWLWWLWZ22}
Chen, C., Zhou, J., Zheng, L., Wu, H., Lyu, L., Wu, J., Wu, B., Liu, Z., Wang,
  L., Zheng, X.: Vertically federated graph neural network for
  privacy-preserving node classification. In: Proceedings of the Thirty-First
  International Joint Conference on Artificial Intelligence, {IJCAI} 2022,
  Vienna, Austria, 23-29 July 2022. pp. 1959--1965. ijcai.org (2022)

\bibitem{medical_dignoise}
FeatureCloud: Transforming health care and medical research with federated
  learning (2020)

\bibitem{fu2022label}
Fu, C., Zhang, X., Ji, S., Chen, J., Wu, J., Guo, S., Zhou, J., Liu, A.X.,
  Wang, T.: Label inference attacks against vertical federated learning. In:
  31st USENIX Security Symposium (USENIX Security 22), Boston, MA (2022)

\bibitem{Fu2021VF2BoostVF}
Fu, F., Shao, Y., Yu, L., Jiang, J., Xue, H., Tao, Y., Cui, B.: Vf2boost: Very
  fast vertical federated gradient boosting for cross-enterprise learning.
  Proceedings of the 2021 International Conference on Management of Data
  (2021)

\bibitem{gu2017badnets}
Gu, T., Dolan-Gavitt, B., Garg, S.: Badnets: Identifying vulnerabilities in the
  machine learning model supply chain. arXiv: Cryptography and Security  (2017)

\bibitem{he2016deep}
He, k., Zhang, x., Ren, s., sun, j.: Deep residual learning for image
  recognition. CVPR  (2016)

\bibitem{model_inversion}
He, Z., Zhang, T., Lee, R.B.: Model inversion attacks against collaborative
  inference. In: Proceedings of the 35th Annual Computer Security Applications
  Conference. p. 148–162 (2019)

\bibitem{hitaj2017deep}
Hitaj, B., Ateniese, G., Pérez-Cruz, F.: Deep models under the gan:
  Information leakage from collaborative deep learning. The ACM Conference on
  Computer and Communications Security, (CCS) pp. 603--618 (2017)

\bibitem{1997Long}
Hochreiter, S., Schmidhuber, J.: Long short-term memory. Neural Computation
  \textbf{9}(8),  1735--1780 (1997)

\bibitem{Hu2019FDMLAC}
Hu, Y., Niu, D., Yang, J., Zhou, S.: Fdml: A collaborative machine learning
  framework for distributed features. Proceedings of the 25th ACM SIGKDD
  International Conference on Knowledge Discovery \& Data Mining  (2019)

\bibitem{jin2021cafe}
Jin, X., Chen, P.Y., Hsu, C.Y., Yu, C.M., Chen, T.: Cafe: Catastrophic data
  leakage in vertical federated learning. Advances in Neural Information
  Processing Systems  \textbf{34},  994--1006 (2021)

\bibitem{kairouz2021advances}
Kairouz, P., McMahan, H.B., Avent, B., Bellet, A., Bennis, M., Bhagoji, A.N.,
  Bonawitz, K., Charles, Z., Cormode, G., Cummings, R., et~al.: Advances and
  open problems in federated learning. Foundations and Trends{\textregistered}
  in Machine Learning  \textbf{14}(1--2),  1--210 (2021)

\bibitem{Kaissis2020SecurePA}
Kaissis, G., Makowski, M.R., R{\"u}ckert, D., Braren, R.F.: Secure,
  privacy-preserving and federated machine learning in medical imaging. Nature
  Machine Intelligence  \textbf{2},  305--311 (2020)

\bibitem{krizhevsky2009learning}
Krizhevsky, A., Hinton, G., et~al.: Learning multiple layers of features from
  tiny images  (2009)

\bibitem{yujun2017}
Lin, Y., Han, S., Mao, H., Wang, Y., Dally, W.J.: Deep gradient compression:
  Reducing the communication bandwidth for distributed training  (2017)

\bibitem{liu2020backdoor}
Liu, Y., Yi, Z., Chen, T.: Backdoor attacks and defenses in feature-partitioned
  collaborative learning. arXiv preprint arXiv:2007.03608  (2020)

\bibitem{Liu2018TrojaningAO}
Liu, Y., Ma, S., Aafer, Y., Lee, W.C., Zhai, J., Wang, W., Zhang, X.: Trojaning
  attack on neural networks. In: Network and Distributed System Security
  Symposium, {NDSS} (2018)

\bibitem{maas2011learning}
Maas, A., Daly, R.E., Pham, P.T., Huang, D., Ng, A.Y., Potts, C.: Learning word
  vectors for sentiment analysis. In: Proceedings of the 49th annual meeting of
  the association for computational linguistics: Human language technologies.
  pp. 142--150 (2011)

\bibitem{conf/uss/NguyenRCYMFMMMZ22}
Nguyen, T.D., Rieger, P., Chen, H., Yalame, H., M{\"{o}}llering, H.,
  Fereidooni, H., Marchal, S., Miettinen, M., Mirhoseini, A., Zeitouni, S.,
  Koushanfar, F., Sadeghi, A., Schneider, T.: {FLAME:} taming backdoors in
  federated learning. In: Butler, K.R.B., Thomas, K. (eds.) 31st {USENIX}
  Security Symposium, {USENIX} Security 2022, Boston, MA, USA, August 10-12,
  2022. pp. 1415--1432. {USENIX} Association (2022)

\bibitem{ozdayi2021defending}
Ozdayi, S.M., Kantarcioglu, M., Gel, R.Y.: Defending against backdoors in
  federated learning with robust learning rate. THIRTY-FIFTH AAAI CONFERENCE ON
  Artificial Intelligence, {AAAI} pp. 9268--9276 (2021)

\bibitem{BHI_dataset}
P.Mooney: Breast histopathology images (2017)

\bibitem{conf/ndss/RiegerNMS22}
Rieger, P., Nguyen, T.D., Miettinen, M., Sadeghi, A.: Deepsight: Mitigating
  backdoor attacks in federated learning through deep model inspection. In:
  29th Annual Network and Distributed System Security Symposium, {NDSS} 2022,
  San Diego, California, USA, April 24-28, 2022. The Internet Society (2022)

\bibitem{russakovsky2015imagenet}
Russakovsky, O., Deng, J., Su, H., Krause, J., Satheesh, S., Ma, S., Huang, Z.,
  Karpathy, A., Khosla, A., Bernstein, M., et~al.: Imagenet large scale visual
  recognition challenge. International journal of computer vision
  \textbf{115}(3),  211--252 (2015)

\bibitem{shafahi2018poison}
Shafahi, A., Huang, R.W., Najibi, M., Suciu, O., Studer, C., Dumitras, T.,
  Goldstein, T.: Poison frogs! targeted clean-label poisoning attacks on neural
  networks. NIPS 2018  (2018)

\bibitem{conf/ndss/ShejwalkarH21}
Shejwalkar, V., Houmansadr, A.: Manipulating the byzantine: Optimizing model
  poisoning attacks and defenses for federated learning. In: 28th Annual
  Network and Distributed System Security Symposium, {NDSS} 2021, virtually,
  February 21-25, 2021. The Internet Society (2021)

\bibitem{simonyan2015very}
Simonyan, K., Zisserman, A.: Very deep convolutional networks for large-scale
  image recognition. international conference on learning representations
  (2015)

\bibitem{credit_risk}
Webank: Utilization of fate in risk management of credit in small and micro
  enterprises (2018)

\bibitem{xie2020dba}
Xie, C., Huang, K., Chen, P.Y., Li, B.: Dba: Distributed backdoor attacks
  against federated learning. ICLR  (2020)

\bibitem{conf/acsac/XuWKLP22}
Xu, J., Wang, R., Koffas, S., Liang, K., Picek, S.: More is better (mostly): On
  the backdoor attacks in federated graph neural networks. In: Annual Computer
  Security Applications Conference, {ACSAC} 2022, Austin, TX, USA, December
  5-9, 2022. pp. 684--698. {ACM} (2022)

\bibitem{yang2019federated}
Yang, Q., Liu, Y., Chen, T., Tong, Y.: Federated machine learning: Concept and
  applications. ACM Transactions on Intelligent Systems and Technology (TIST)
  \textbf{10}(2),  1--19 (2019)

\bibitem{Zhang2021SecureBA}
Zhang, Q., Gu, B., Deng, C., Huang, H.: Secure bilevel asynchronous vertical
  federated learning with backward updating. In: AAAI (2021)

\bibitem{conf/icml/ZhangPSYMMR022}
Zhang, Z., Panda, A., Song, L., Yang, Y., Mahoney, M.W., Mittal, P.,
  Ramchandran, K., Gonzalez, J.: Neurotoxin: Durable backdoors in federated
  learning. In: International Conference on Machine Learning, {ICML} 2022,
  17-23 July 2022, Baltimore, Maryland, {USA}. Proceedings of Machine Learning
  Research, vol.~162, pp. 26429--26446. {PMLR} (2022)

\bibitem{NEURIPS2019_60a6c400}
Zhu, L., Liu, Z., Han, S.: Deep leakage from gradients. In: Advances in Neural
  Information Processing Systems. vol.~32. Curran Associates, Inc. (2019)

\end{thebibliography}

\end{document}